\documentclass[twocolumn,secnumarabic,amssymb, nobibnotes, aps, prl, superscriptaddress]{revtex4-2}

\usepackage{graphicx}
\usepackage{dcolumn}
\usepackage{bm}
\usepackage{amsmath}
\usepackage{enumerate}
\usepackage{setspace}
\usepackage{dsfont}
\usepackage{subfigure}
\usepackage{multirow}
\usepackage{indentfirst} 
\usepackage {mathrsfs}
\usepackage{color}

\usepackage[utf8]{inputenc}
\usepackage[T1]{fontenc}
\usepackage{mathptmx}
\usepackage{etoolbox}
\usepackage{upgreek}
\usepackage{subfigure}

\usepackage{threeparttable}  
\usepackage[pdfstartview=FitH,
CJKbookmarks=true,
bookmarksnumbered=true,
bookmarksopen=true,
colorlinks, 
pdfborder=001,
linkcolor=blue,
anchorcolor=blue,
citecolor=blue
]{hyperref}

\begin{document}

\title{Hundred-Femtosecond-Level Concise Optical Time Delay Measurement System Based on Linear Optical Sampling}%

\author{Yufei Zhang}%
\affiliation{State Key Laboratory of Advanced Optical Communication Systems and Networks, School of Electronics, and Center for Quantum Information Technology, Peking University, Beijing 100871, China}

\author{Ziyang Chen}
\altaffiliation{The authors to whom correspondence may be addressed:\\ chenziyang@pku.edu.cn, hongguo@pku.edu.cn.}
\affiliation{State Key Laboratory of Advanced Optical Communication Systems and Networks, School of Electronics, and Center for Quantum Information Technology, Peking University, Beijing 100871, China}

\author{Dongrui Yu}
\affiliation{State Key Laboratory of Advanced Optical Communication Systems and Networks, School of Electronics, and Center for Quantum Information Technology, Peking University, Beijing 100871, China}

\author{Jialin Niu}
\affiliation{State Key Laboratory of Information Photonics and Optical Communications, Beijing University of Posts and Telecommunications, Beijing 100876, China}

\author{Xing Chen}
\affiliation{State Key Laboratory of Information Photonics and Optical Communications, Beijing University of Posts and Telecommunications, Beijing 100876, China}

\author{Hong Guo}
\altaffiliation{The authors to whom correspondence may be addressed:\\ chenziyang@pku.edu.cn, hongguo@pku.edu.cn.}
\affiliation{State Key Laboratory of Advanced Optical Communication Systems and Networks, School of Electronics, and Center for Quantum Information Technology, Peking University, Beijing 100871, China}

\date{\today}%

\begin{abstract}
Fiber-delay measurement is one of the key fundamental technologies in numerous fields. Here we propose and experimentally demonstrate a high-precision and concise optical time delay measurement system based on the technique of linear optical sampling, reaching the precision better than 100 fs under averaging. The use of only two optical frequency combs without locking the carrier-envelope-offset frequency greatly simplifies the structure of the time-delay measurement system. We also experimentally investigate the current limitations on the precision of the system. The timing jitter noises of two sources are mainly non-common mode, and are both restricted to the frequency sources. Our results indicate that the proposed device can measure fiber length fluctuations below 10 $\upmu\text{m}$, paving the way for further analyses of the external disturbances on the fiber link.
\end{abstract}

\maketitle

\section{Introduction}
Optical fiber has extremely high sensitivity to external physical disturbances such as temperature, strain, pressure, etc.\cite{RN612} Therefore, optical fiber sensors have important applications in many fields, specifically, monitoring of aeroplanes, oil and gas industries, railway infrastructure, geologic $\text{CO}_2$ sequestration, and sensing of temperature vibration, structural health of infrastructure etc.\cite{RN616, RN617, RN618, RN632, RN633}

In optical fiber sensors, how to measure the length of optical fiber with high precision and the length fluctuations caused by external disturbances is a basic and significant problem. In fact, measuring the length of the optical fiber and measuring the propagation delay of the optical pulse in the optical fiber can be transformed into each other. At present, the widely used optical fiber length measurement methods mainly include optical time domain reflectometry (OTDR) and optical frequency domain reflectometry (OFDR).\cite{RN612} The measurement is taken by coupling the optical pulse at one end of the fiber under test (FUT) and receiving the Rayleigh, Brillouin or Raman scattering signal generated by its propagation, and measuring the optical time delay $t$ corresponding to the two events in the time domain or frequency domain. Measurement of fiber length yields $z=\frac{v_\text{g}}{2}t$, where $v_\text{g}$ is the group velocity propagating in the fiber. Its measurement accuracy $\Delta z=v_\text{g}\frac{\tau}{2}$ mainly depends on the pulse width $\tau$, which is generally on the order of millimeters to meters.\cite{RN612} Another solution for measuring the length of optical fiber is realized by using a frequency-shifted asymmetric Sagnac interferometer.\cite{RN611} The basic idea is that light of different frequencies will produce different phase delays when propagating in the same section of optical fiber, and this phase delay difference containing information about the length of the optical fiber can be measured by an interferometer. When using this method to measure $10^5$ m fiber, the accuracy reaches the highest value $10^{-6}$, corresponding to an absolute resolution of 0.1 m. Furthermore, researchers achieved $\pm1$ ps accuracy by measuring the peak-to-peak value of fiber delay fluctuation.\cite{RN614}

However, although these existing schemes can basically meet the general industrial applications, their devices are still relatively complex and have limited accuracy, which are still insufficient for higher-precision scientific applications such as geodesy,\cite{RN6, RN519, RN40, RN525} navigation\cite{RN8, article} and basic sciences.\cite{RN40, RN7, RN114} In this paper, we propose a concise optical fiber time delay measurement device, whose basic idea is to magnify the tiny time interval by the method of linear optical sampling (LOS)\cite{RN521, RN21, RN14, RN17, RN46, RN576, Abu} for measurement. Our system uses two mode-locked lasers with a repetition frequency difference $\Delta f_\text{r}$, and the pulse train of one laser beats with the pulse train of the other laser after passing through the FUT. Through the LOS process in the time domain and post-processing methods such as Hilbert filtering and Gaussian fitting, the magnification of the tiny delay in the time domain is realized with the factor $\frac{f_\text{r}+\Delta f_\text{r}}{\Delta f_\text{r}}\approx10^5$. In the experiment, we use an adjustable fiber delay line as the FUT. Under the delay step value of 10 ps, the standard error of mean of single-point delay measurements reaches an accuracy of less than 100 fs. That is to say, the new concise device we propose realizes the measurement of the fiber delay with an accuracy of 100 fs, corresponding to the fiber length measurement accuracy of $10^{-5}$ m. This means that the proposed device can measure fiber length fluctuations on the order of 10 $\upmu\text{m}$, and further analyze the impact of external environment changes on the fiber link. We believe that this concise device has the prospect of further integration and miniaturization, and its potential advantages of high measurement accuracy are expected to open a new pathway in the fields of optical fiber sensing and high-precision metrology.
\begin{figure*}[t]
	\centering
	\includegraphics[width= 0.7 \linewidth]{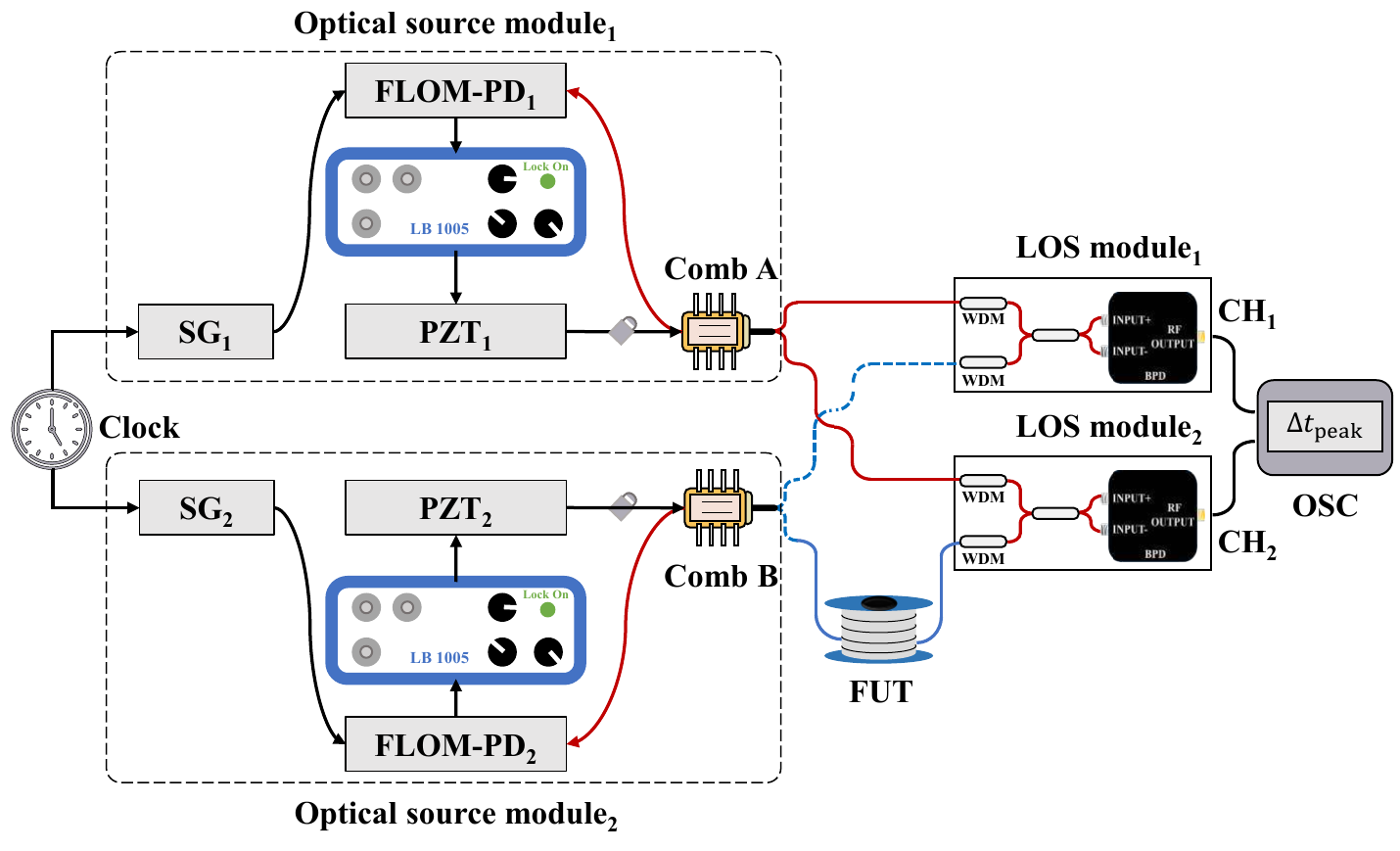}
	\caption{Schematic of the experimental setup. Optical source modules, LOS modules and a rubidium clock which is used to synchronize the two signal generators (SGs) are included. A single optical source module is mainly comprised of a signal generator and an optical frequency comb. Moreover, a fiber loop optical microwave-phase detector (FLOM-PD), a commercial high-precision feedback control device (New Focus LB1005) and a piezoelectric ceramics (PZT) driver are used for high-precision repetition frequency locking of OFC. The repetition frequencies for comb A and comb B are $f_\text{r}+\Delta f_\text{r}=250~\text{MHz}+2.5~\text{kHz}$ and $f_\text{r}=250~\text{MHz}$ respectively. The LOS module$_1$ is used as the reference and the LOS module$_2$ is used to obtain the time information of the fiber under test (FUT). A single LOS module contains two wavelength division multiplexers (WDMs), followed by a beam splitter and a balanced photodetector. OSC: Oscilloscope.}
	\label{System}
\end{figure*}

In this paper, we first describe the high precision optical time delay measurement, within which the experimental setup and the data processing are introduced. Then we present the main results and discussions. A summary of the paper is given at the end. As for the principles of optical time delay measurement based on LOS, we include them in the appendix.

\section{High Precision Optical Time Delay Measurement}
In this section, we give a detailed introduction of our proposed optical time delay measurement system.
\subsection{Experimental Setup}
Here, we propose a concise scheme for high-precision optical time delay measurement with only two optical frequency combs (OFCs). We adopt an adjustable electronically controlled optical delay line (LightSource LT-ODL-15-120-1-FA, a specific product of a Chinese company with nominal precision 10.16 fs) as FUT to simulate the small time delay $\tau$. Exploiting it, the pulses generated by the optical comb B is delayed $\tau$. Our system has the ability to measure such small time delay with high precision. 

Besides FUT, the schematic of our system in Fig. \ref{System} also includes optical source modules, LOS modules and a rubidium clock (1E-12@1 s) whose output 10 MHz frequency signal is used to synchronize the two signal generators. A single optical source module is mainly comprised of a signal generator (Rigol DSG821) and a nonlinear polarization rotation (NPR) mode-locked OFC (homemade products at the wavelength of 1550 nm, with output power 26.97 mW for comb A, and 37.42 mW for comb B). Moreover, a fiber loop optical microwave-phase detector (FLOM-PD), a commercial high-precision feedback control device (New Focus LB1005) and a piezoelectric ceramics (PZT) driver also act as auxiliary parts in a single optical source module. These are used for high-precision repetition frequency locking of OFC, whose accuracy is extremely important for the signal quality obtained by LOS. 

In the process of frequency locking, phase detection is one of the core techniques. Traditionally, an electrical phase detector in a phase-locked loop is used to realize the phase detection of the reference signal and the signal to be measured. However, this scheme inevitably introduces noise in the photoelectric conversion process. In this experiment, we use FLOM-PDs for their higher phase detection sensitivity and less affection by the laser intensity noise and environmental changes.\cite{RN123} Thus experimentally, to lock the repetition frequency $f_\text{r}=250~\text{MHz}$ (for comb B, or equally $f_\text{r}+\Delta f_\text{r}=250~\text{MHz}+2.5~\text{kHz}$ for comb A) with the signal generator, first the output of OFC is split into two beams. One of them is used for phase detection by FLOM-PD to lock the repetition frequency, while the other is used for the following experiment. Second, the error signal obtained in the phase detection is input into LB1005. The proportional and integral parameters are adjusted, and the feedback voltage signal is input into the PZT driver. The repetition frequency of OFC is adjusted by controlling the cavity length through the expansion and contraction of PZT.

\begin{figure}[htbp]
    \centering
    \includegraphics[width=8.5cm]{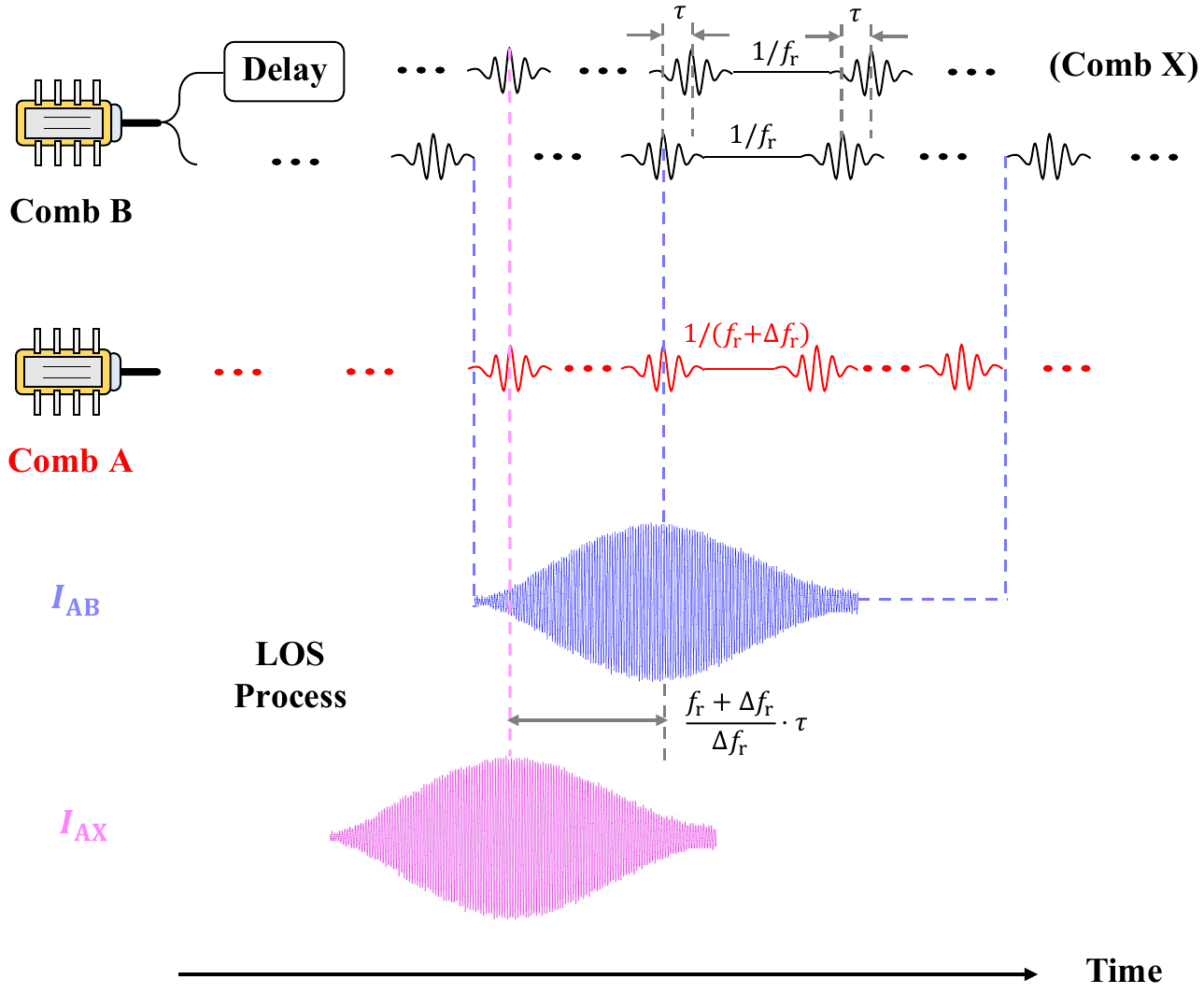}
    \caption{Schematic of the principle of the small optical time delay measurement system based on LOS. The small time delay $\tau$ introduced in the pulse sequence is magnified to $\frac{f_\text{r}+\Delta f_\text{r}}{\Delta f_\text{r}}\cdot\tau$ time shift of the LOS signal pattern.}
    \label{LOS}
\end{figure}

The LOS module contains two subparts, where one is the LOS between the optical comb A and the optical comb B without delay, and the other is the LOS between the optical comb A and the optical comb B after the delay. The two signals enter the 50:50 2*2 beam splitter after passing through wavelength division multiplexers (WDMs). The two outputs of the beam splitter respectively enter the two inputs of the balanced photodetector (Thorlabs PDB450C). The simplfied complex light fields of the two OFCs $A_\text{1}=\sum_n E_ne^{\text{i}(\omega_nt+\phi_n)}$ and $A_\text{2}=\sum_m E_me^{\text{i}(\omega_mt+\phi_m)}$ satisfy the transformation after passing through the 2*2 beam splitter in LOS $\text{module}_1$\cite{RN634}
\begin{equation}
    \begin{pmatrix}
    A_\text{1}^{\prime} \\ A_\text{2}^{\prime}
    \end{pmatrix}
    =
    \begin{pmatrix}
    \frac{\sqrt{2}}{2} & \frac{\sqrt{2}}{2}\text{i} \\ \frac{\sqrt{2}}{2}\text{i} & \frac{\sqrt{2}}{2}
    \end{pmatrix}
    \begin{pmatrix}
    A_\text{1}\\A_\text{2}
    \end{pmatrix},
    \label{splitter}
\end{equation}
and the output current after passing the balanced photodetector satisfies
\begin{equation}
    I_\text{out1}=I_\text{combA}-I_\text{combB}\propto|A_\text{1}^{\prime}|^2-|A_\text{2}^{\prime}|^2.
    \label{photodetector}
\end{equation}
Substitute Eq.(\ref{splitter}) into Eq. (\ref{photodetector}), we can get
\begin{equation}
    \begin{aligned}
    I_\text{out1}&\propto2\Im(A_\text{1}A_{2}^*)\\
    &=2\sum_{mn}E_nE_m\sin[(\omega_n-\omega_m)t+(\phi_n-\phi_m)],
    \label{bpd}
    \end{aligned}
\end{equation}
where $n$ ($m$) refers to the number of the cavity modes of OFC, with $\omega_n$ ($\omega_m$) and $\phi_n$ ($\phi_m$) denoting the corresponding frequency and phase of each cavity mode. Since the two optical combs are actually locked on the same rubidium clock, $\Delta\phi=\phi_n-\phi_m$ here can be regarded as time-independent. We also notice that the spectra of the OFCs we use do overlap. Therefore, the fundamental frequency of the signal obtained by LOS is the difference between the fundamental frequencies of the two optical combs, which is $\Delta f_\text{r}$. Here, the outputs of the balanced detectors are respectively connected to the channels 1 and 2 of the oscilloscope, which are two LOS signals. From the LOS procedure\cite{RN21} shown in Fig. \ref{LOS}, the small time delay $\tau$ introduced in the pulse sequence is magnified to $\frac{f_\text{r}+\Delta f_\text{r}}{\Delta f_\text{r}}\cdot\tau$ in the LOS signal pattern, thereby achieving higher precision time delay measurement. And thus, we finally derive the optical time delay $\tau$ from
\begin{equation}
    \tau = \frac{\Delta f_\text{r}}{f_\text{r}+\Delta f_\text{r}}\cdot\Delta t_\text{peak},
    \label{time delay}
\end{equation}
whose detailed derivation is presented in the appendix.

Eq. (\ref{time delay}) indicates that with the design of two subparts of the LOS module, the small optical time delay $\tau$ is magnified to the relative delay of LOS patterns between two channels of the oscilloscope, denoted by $\Delta t_\text{peak}$, which is easy to measure. In our experiment, the nominal optical time delay of the delay line is changed step by step from 0 to 60 ps with an interval of 10 ps. At each nominal value, we measure 100 sets of data (one set of data includes the data of channels 1 and 2 of the oscilloscope). We set the sampling rate of the oscilloscope to 500 MHz, and the amount of data per channel of each set of data is 100 k points, corresponding to 200 $\upmu$s sampling time. Here we point out that in order to acquire the full waveforms of these two LOS signals, by controlling the length of fiber experimentally, we set the two waveforms close to each other when the delay line is 0 ps. Thus, even when the delay line reaches its maximum limit in our experiment (60 ps), the two waveforms only separates about 6 $\upmu$s, indicating that the sampling time of 200 $\upmu$s we use is long enough to acquire the full waveforms.

\subsection{Data Processing}
\begin{figure}[htbp]
    \centering
    \includegraphics[width =0.9 \linewidth]{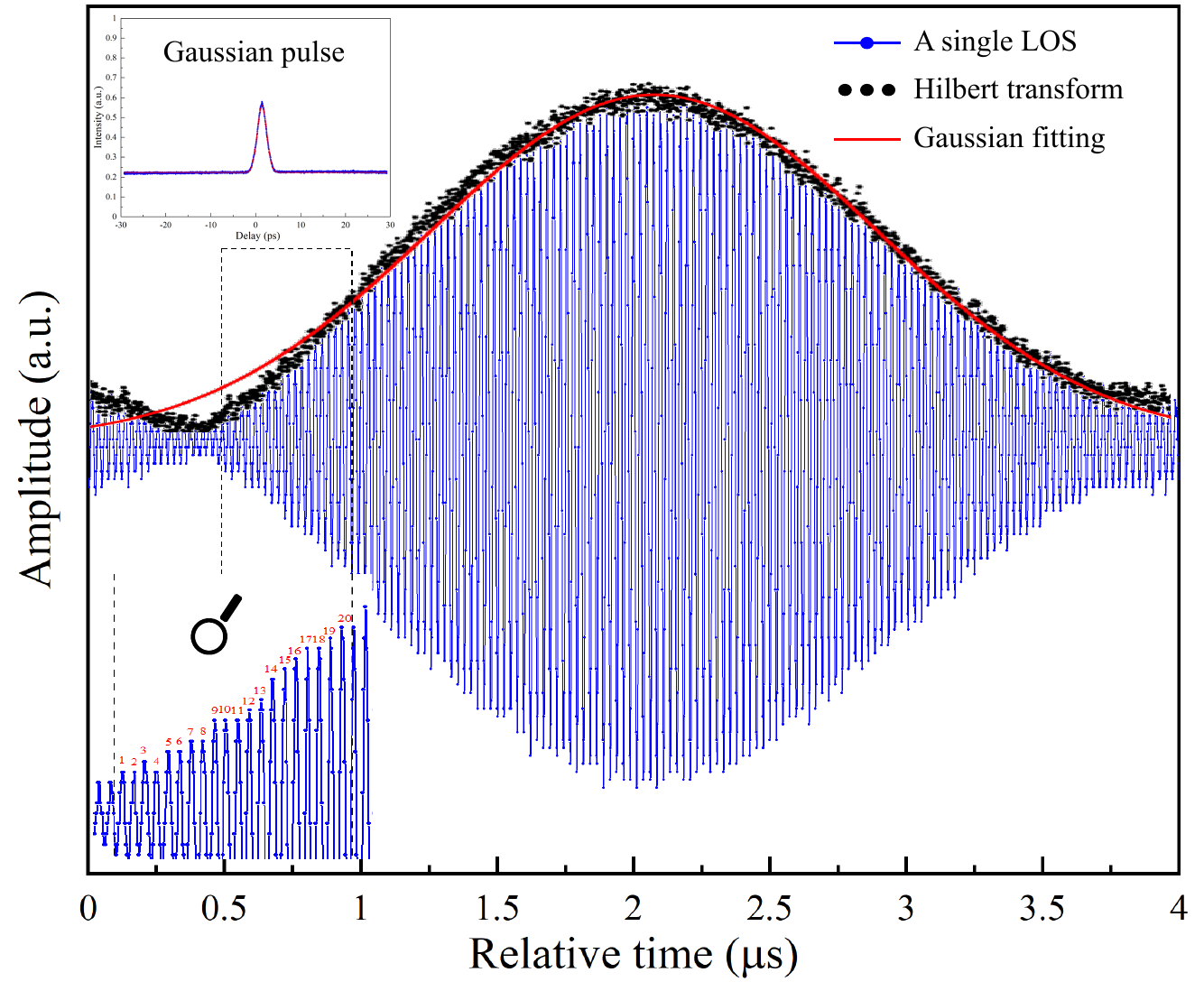}
    \caption{A single pulse waveform of the acquired LOS signal, whose waveform is still Gaussian shaped (blue). The effect of Hilbert transform with taking the absolute value (black) and Gaussian fitting (red) are also manifested here. In the top left corner, the autocorrelator shows the optical pulse before entering the LOS process is Gaussian shaped. In the bottom left corner, we zoom in the period of 0.5 to 1 $\upmu$s, making it clearer to see an oscillation near 45 MHz, which results from the bandwidth of the balanced photodetector.}
    \label{Fitting}
\end{figure}

The blue line in Fig. \ref{Fitting} shows a single pulse in the acquired LOS signal. Here we notice that the oscillation in the signal is a consequence of the 45 MHz bandwidth of the balanced photodetector. Moreover, according to the dispersion in the cavity of the NPR mode-locked laser, the output pulse is Gaussian shaped. The small figure in the top left corner of Fig. \ref{Fitting} shows the waveform of the signal before the LOS procedure, acquired from the autocorrelator. Thus, based on Eq. (\ref{bpd}), the LOS signal does not change the waveform of the original pulse signal, and we use Gaussian fitting to find the peak. However, it can also be seen from Eq. (\ref{bpd}) that the LOS signal collected in the experiment contains phase information. In order to easily extract the envelope information, the Hilbert transform is used to separate the phase and the amplitude terms. After Hilbert transforming the LOS signal and taking the absolute value of the result, the envelope of the LOS signal is easily extracted. Next, by performing Gaussian fitting on the envelope and recording the central time corresponding to the peak, the fitting of a single LOS signal is completed. Fig. \ref{Fitting} shows the fitting result.

The fitting accuracy $\sigma_\text{F}$ mainly depends on the error of the central time corresponding to the peak during Gaussian fitting. In this experiment, the fitting accuracy of the LOS pattern is below 5.0 ns, resulting in $\sigma_\text{F}<5.0~\text{ns}\cdot\frac{\Delta f_\text{r}}{f_\text{r}+\Delta f_\text{r}}=50~\text{fs}$ according to Eq. (\ref{time delay}). We see from the following analysis that this part of the error has little influence on the accuracy of the final time delay measurement.

\section{Results and Discussions}
We measure 100 sets of data at each nominal delay. After processing the LOS signal to get the time of the peak, we use Eq. (\ref{time delay}) to calculate 100 measured time delay by dividing the amplification factor. We also calculate the $\sigma_\text{SEM}$ (standard error of mean) of these 100 data. The formula for calculating the standard error of mean is
\begin{equation}
    \sigma_\text{SEM}=\frac{\sigma_\text{SD}}{\sqrt{N}},
    \label{SEM}
\end{equation}
where $N=100$ is the amount of data, and $\sigma_\text{SD}$ represents the standard deviation of these 100 statistical data. Since the standard error of mean represents the deviation of the average value of 100 data from the true expected value, $\sigma_\text{SEM}$ is a sound index for evaluating the precision our system can achieve under a certain number of optical time delay measurements.

\begin{figure}[htbp]
    \centering
    \includegraphics[width=8.5cm]{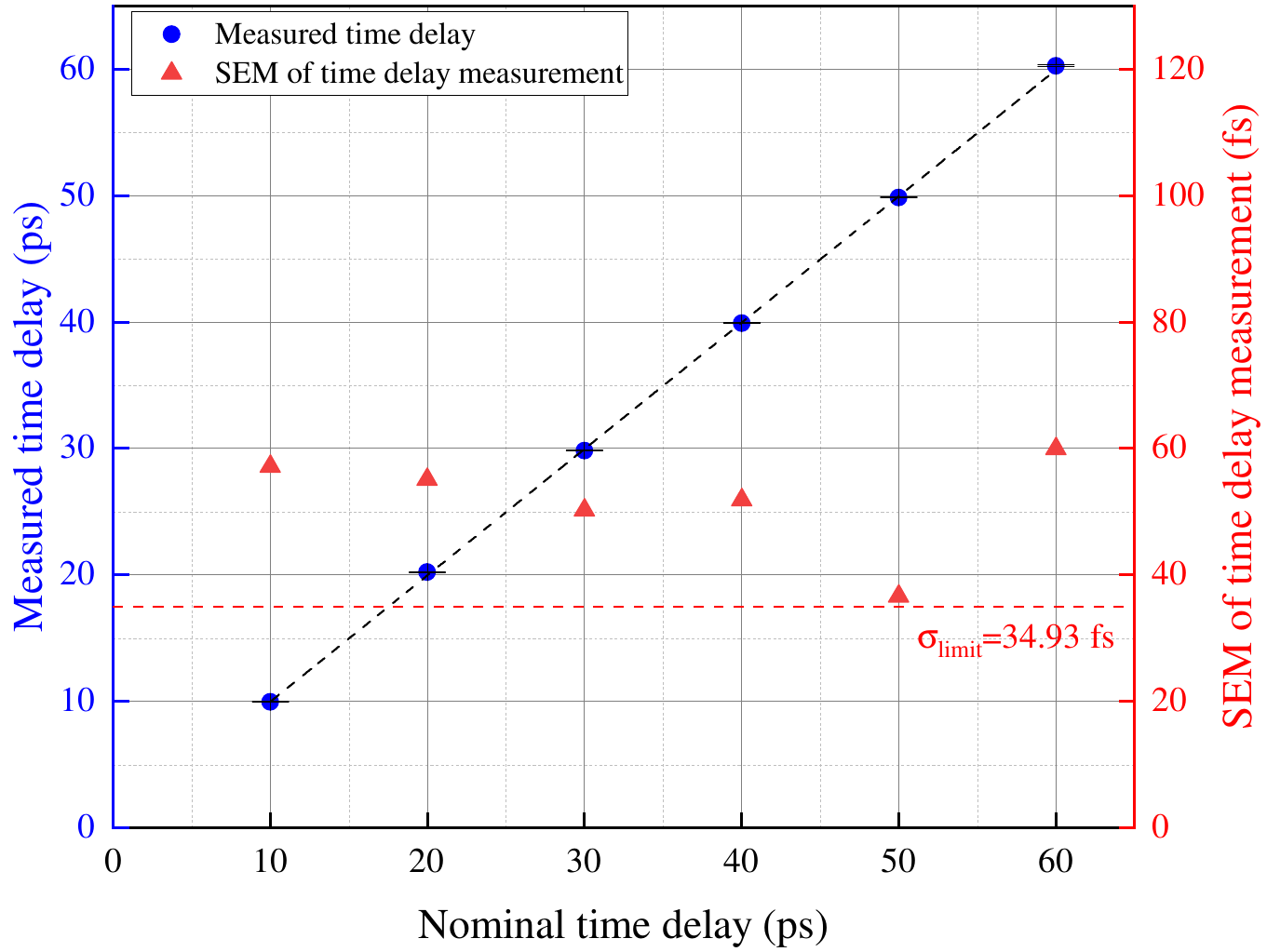}
    \caption{Experimentally measured time delay versus nominal time delay (blue circle). Each point is averaged from 100 samples with error bar equals the standard deviation, and the linearity is $\text{r}=0.99995$. The experimentally measured standard error of mean (SEM) under each time delay is shown in red triangle. The SEM shows the precision our system can achieve in optical time delay measurement, which is below the level of 100 fs. The red dashed line gives the limitation of the accuracy.}
    \label{Result}
\end{figure}

The experiment results are shown in Fig. \ref{Result}. The standard error of mean of our dual-comb time delay measurement system's results are on the magnitude of below 100 fs under every nominal time delay, indicating that we are able to measure the time delay with the precision better than 100 fs.

Here, the stability of the LOS signal is the dominant limitation of the accuracy of our time delay measurement system. Since the physical process of LOS is the interference of dual-comb pulse signals, the timing jitter of the obtained LOS signal should mainly depend on the relative timing jitter of the two optical combs. As the proof, two 250 MHz optical comb signals are respectively connected to the two inputs of the phase noise analyzer (ROHDE \& SCHWARZ FSUP), and the \textit{Double DUT Measurement} mode was selected to measure the relative phase noise of the two signals. To calculate the timing jitter $\sigma_\text{t}$, we refer to the formula\cite{RN120,RN124}
\begin{equation}
    \sigma_\text{t} = \frac{1}{2\uppi f_\text{r}}\sqrt{2\int_{f_\text{L}}^{f_\text{H}}L(f)\text{d}f},
\end{equation}
where $f_\text{r}=250~\text{MHz}$ is the repetition frequency, $f_\text{L}=1~\text{kHz}$ and $f_\text{H}=10~\text{MHz}$ are the lower and upper limits of the integral respectively, $L(f)$ represents the single sideband (SSB) phase noise spectrum, and reaches noise floor beyond $f_\text{H}$. The result of the integral shows that the relative timing jitter is $\sigma_{t_\text{r}}=345.72~\text{fs}$. Thus, the accuracy of our system satisfies 
\begin{equation}
\sigma_\text{SD}^2\ge\sigma_{t_\text{r}}^2+\sigma_\text{F}^2.
\label{limitation}
\end{equation}
We find that $\sigma_\text{SD}\ge349.3~\text{fs}\approx\sigma_{t_\text{r}}$, indicating that the fitting accuracy is the nondominant part. Combined with Eq. (\ref{SEM}), we find $\sigma_\text{SEM}\ge34.93~\text{fs}$, giving the ceiling of the time delay measurement accuracy.
Furthermore, we measure the timing jitters of the two combs to be $\sigma_{t_\text{1}}=254.90~\text{fs}$ and $\sigma_{t_\text{2}}=325.74~\text{fs}$ respectively. They satisfy $\sigma_{t_\text{r}}^2\approx\sigma_{t_1}^2+\sigma_{t_2}^2$ approximately, indicating that the timing jitter noise between the two combs is mainly non-common mode. Thus, the final timing jitter of the LOS signal highly depends on the respective timing jitters of the two combs.

Moreover, since the repetition frequency is locked directly on the signal generator, we also compare the phase noise of the signal generator and the comb. The integral of the timing jitter of the signal generator (frequency source), also from 1 kHz to 10 MHz yields the value of 245.55 fs. Compared with the timing jitter of the two combs (254.90 fs and 325.74 fs), we show that the precision of the FLOM-PD based frequency locking is high enough. So the first further optimization is the improvement of the frequency source. Second, as the analysis above shows, the timing jitters of the two combs have little common-mode noise to offset during the LOS procedure. Thus, the second further optimization is the improvement of the synchronization, where a better clock may be needed. Finally, the improvements of the data processing algorithm would be important when the error term $\sigma_\text{F}$ dominates, and we consider applying maximum likelihood estimation in the future work to locate signal position,\cite{RN644} trying to achieve even higher resolution.

\section{Conclusions}
We design and experimentally realize a concise high-precision optical time delay measurement system based on the method of comb-based LOS. We experimentally use a fiber delay line as the FUT, and adjust the it step by step from 0 to 60 ps with the interval of 10 ps. The measurement result derived from our system results in the standard error of mean below the level of 100 fs. We also investigate the limitations on the precision of our time delay measurement system in detail. The experiment results illustrate that the relative phase noise of the two OFCs constitutes the dominant part of the timing jitter of the LOS signal, which represents the precision our system can reach. Meanwhile, the relative timing jitter integrated by the relative phase noise spectrum is estimated to be the square root of the sum of the two combs' own timing jitters, indicating that the non-common mode noise dominates. Finally, the timing jitters of the respective OFC are also found to be restricted to the performance of the current signal source. Thus, better atomic clocks and signal generators are needed for even higher precision optical time delay measurement. Our experiment results indicates that we are capable of measuring the tiny optical time delay with the precision better than 100 fs, making it possible for the detection of fiber length fluctuations below the level of $10^{-5}$ m. This high precision will surely find its applications in the field of fiber sensors to monitor and detect the external physical disturbances on the fiber length. 

\appendix*
\section{APPENDIX: Principles of LOS Based Optical Time Delay Measurement}
As shown in Fig. \ref{LOS}, the repetition frequency of the two OFCs are $f_\text{r}$ and $f_\text{r}+\Delta f_\text{r}$. Now we consider the principle of the LOS procedure more detailedly.\cite{RN521, RN3, RN122} For brevity, the comb B after delay is marked as comb X. We then denote the interference pattern of comb A and B after LOS as $I_\text{AB}$, and $p_\text{AB}$ denotes the peak number in this interference pattern, corresponding to time $t_{p_\text{AB}}$. On the other hand, each peak of $I_\text{AB}$ corresponds to the "alignment" of the two peaks of A and B at the same moment, indicating the constructive interference. Therefore, we have the following two aspects of analyses.

First, for the peak $p_\text{AB}$, it is generated from the superposition of the $k_\text{BB}^\text{th}$ pulse of the comb B and the $k_\text{AB}^\text{th}$ pulse of the comb A. They can be expressed as
\begin{equation}
    k_\text{BB}=f_\text{r}\cdot t_{p_\text{AB}},
\end{equation}
and
\begin{equation}
    k_\text{AB}=(f_\text{r}+\Delta f_\text{r})\cdot(t_{p_\text{AB}}+\Delta_\text{AB}),
\end{equation}
where $\Delta_\text{AB}$ represents the absolute time difference between A and B. Second, if we start timing from one specific peak in $I_\text{AB}$, and mark it as the zeroth peak, then the $\text{i}^\text{th}$ peak after it corresponds to the case where the peak numbers between comb B and A is i. We derive from this analysis that
\begin{equation}
    p_\text{AB} = k_\text{AB}-k_\text{BB}=f_\text{r}\cdot\Delta_\text{AB}+\Delta f_\text{r}\cdot(t_{p_\text{AB}}+\Delta_\text{AB}),
\end{equation}
indicating that
\begin{equation}
    \Delta_\text{AB}=\frac{p_\text{AB}-\Delta f_\text{r}\cdot t_{p_\text{AB}}}{f_\text{r}+\Delta f_\text{r}}.
    \label{four}
\end{equation}

Similar analyses can be made to the interference pattern from LOS between comb A and X, resulting in
\begin{equation}
    k_\text{XX}=f_\text{r}\cdot t_{p_\text{AX}},
\end{equation}
and 
\begin{equation}
    k_\text{AX}=(f_\text{r}+\Delta f_\text{r})\cdot(t_{p_\text{AX}}+\Delta_\text{AX}).
\end{equation}
Thus we have
\begin{equation}
    p_\text{AX} = k_\text{AX}-k_\text{XX}=f_\text{r}\cdot\Delta_\text{AX}+\Delta f_\text{r}\cdot(t_{p_\text{AX}}+\Delta_\text{AX}),
\end{equation}
and
\begin{equation}
    \Delta_\text{AX} = \frac{p_\text{AX}-\Delta f_\text{r}\cdot t_{p_\text{AX}}}{f_\text{r}+\Delta f_\text{r}}.
    \label{8}
\end{equation}

Combine Eq. ($\ref{four}$) and Eq. ($\ref{8}$), we have
\begin{equation}
\begin{aligned}
    \Delta_\text{BX} &= \Delta_\text{BA}+\Delta_\text{AX}\\
    &= -\Delta_\text{AB}+\Delta_\text{AX}\\
    &= \frac{1}{f_\text{r}+\Delta f_\text{r}}\cdot[p_\text{AX}-p_\text{AB}-\Delta f_\text{r}\cdot(t_{p_\text{AX}}-t_{p_\text{AB}})].
\end{aligned}
\end{equation}

The choice of the peak numbers $p_\text{AX}$ and $p_\text{AB}$ is relative. Notice that here $\Delta_\text{BX}$ corresponds to the small time difference, namely delay $\tau$ in the optical pulse sequence, while $(t_{p_\text{AX}}-t_{p_\text{AB}})$ represents the time shift in the LOS signal pattern. Finally, the measured small time delay $\tau$ is calculated from the peak shift of the LOS signal $\Delta t_\text{peak}$ by
\begin{equation}
    \tau = \frac{\Delta f_\text{r}}{f_\text{r}+\Delta f_\text{r}}\cdot\Delta t_\text{peak}.
    \label{time delay supplementary}
\end{equation}

We theoretically conclude from the above analyses that by introducing a second comb A to perform LOS, we can measure the optical time delay imposed to one of the pulse trains divided from the output of comb B. By measuring the value of $\Delta t_\text{peak}$, we can give the exact value of the time delay $\tau$ directly from Eq. (\ref{time delay supplementary}).

\begin{acknowledgments}
The authors acknowledge the support of Dr. Sheng Li and Dr. Chao Zhou of Peking University, for the valuable and enlightening discussions with them. This work was supported by the National Natural Science Foundation of China (Grants Nos. 62201012, and 61531003), and the National Hi-Tech Research and Development (863) Program.
\end{acknowledgments}

%


\end{document}